\documentclass[twocolumn, showpacs, amsmath, amssymb, amsthm, superscriptaddress, letterpaper]{revtex4}

\usepackage{epsfig,amsfonts}
\usepackage{times}
\usepackage{array}
\usepackage{epstopdf}
\usepackage{graphicx}
\usepackage{mathtools}
\usepackage{dsfont}
\usepackage{xr}

\def\>{\rangle}
\def\<{\langle}

\newcommand{\eq}[1]{Eq.~(\ref{eq:#1})}
\newcommand{\fig}[1]{Fig.~(\ref{fig:#1})}
\newcommand{\secref}[1]{Sec.~\ref{sec:#1}}

\newcommand\idop{\mathds{1}}

\newcommand{\ignore}[1]{}

\newcommand{\ket}[1]{{|}#1{\rangle}}

\begin{document}
\title{Correcting coherent quantum errors by going with the flow}
%\title{Automatic twirling of uncontrolled, correlated qubit rotations in standard quantum error correction}

\author{Wayne M. Witzel}
	\affiliation{Center for Computing Research, Sandia National Laboratories, Albuquerque, New Mexico 87185 USA}
\author{Anand Ganti}
	\affiliation{Cyber Security Initiatives, Sandia National Laboratories, Albuquerque, New Mexico 87185 USA}
\author{Tzvetan S. Metodi}
    \affiliation{Center for Computing Research, Sandia National Laboratories, Albuquerque, New Mexico 87185 USA}
	\affiliation{Current affiliation: Quantinuum, Broomfield, CO 80021, USA}
\date{\today}

\begin{abstract}
The performance of a given quantum error correction (QEC) code depends upon the noise model
that is assumed.  Independent Pauli noise, applied after each quantum
operation, is a simplistic noise model that is easy to simulate and understand
in the context of stabilizer codes.  Although such a noise model is artificial,
it is equivalent to independent, random, unbiased qubit rotations.  What about
spatially or temporally correlated qubit rotations?  Such a noise model 
is applicable to global operations (e.g., NMR or ESR), common control sources 
(e.g., lasers), or slow drift (e.g., charge or magnetic noise) in various qubit technologies.  
In the worst case, such errors can combine constructively and result
in a post-correction failure rate that increases with the number of error correction
cycles.  However, we show that this worst case does not
generally arise unless taking active corrective actions while
performing QEC.  That is, by employing virtual
Pauli frame updates ("passive" error correction) rather than physical corrections
("active" error correction), coherent errors do not compound appreciably.
Starting in a random Pauli frame is also advantageous.  In fact, through perturbation
theory arguments and supporting numerical simulations, we show that the logical
qubit performance beyond distance 3 for correlated single-qubit Hamiltonian noise models (i.e., global errant qubit rotations), 
when employing these ``lazy'' strategies, essentially 
matches the performance of Pauli noise model with the same process fidelity
(fidelity after one application).  In a more general circuit model of noise,
correlations may add constructively within syndrome extraction rounds but 
Pauli frame randomization from passive error correction mitigates this effect across multiple rounds.
\end{abstract}
\maketitle

% ======================== INTRO ============================================== %

\section{Introduction}\label{sec:intro}
There are numerous challenges in realizing a functional, utility-scale quantum computer that could far exceed the performance of its classical counterpart over a range of applications.  A principal challenge is the inevitable decoherence from a noisy environment which, if not mitigated, will reduce a quantum computer to a quasi-random number generator.  The theory of fault tolerant quantum computing provides a constructive proof for accurately simulating arbitrarily large quantum circuits in the presence of noise under certain assumptions.  A key assumption is that noise has limited spatial and temporal correlations.  Once these noise assumptions are satisfied a noise threshold is established.  If the noise strength is below this threshold (an accuracy threshold), arbitrary large quantum circuits can be simulated with an arbitrarily small, but non-zero, error ~\cite{AB96, Knill1998, AGP05, Raussendorf2007}.

In the study of fault tolerant quantum computing, a model of independent, discrete noise is typically used as a convenient proxy for realistic noise.  In this model, quantum gates act as probabilistic devices that either behave perfectly with probability $1-p$ or fail with probability $p$ and become some arbitrary superoperator.  A special case of the independent discrete noise is the independent Pauli noise where a gate behaves as the ideal gate followed by I, X, Y, or Z with respective probabilities.  At the other extreme, models with correlated errors, a noise channel that does not factorize into a product of single-qubit noise channels, are particular pernicious and must meet stringent requirements to exhibit an accuracy threshold~\cite{Aharonov2006, Iverson2020}.  Uncorrelated, coherent errors are an intermediate class and the focus of this work.  Coherent errors may add constructively or destructively in a quantum circuit, scaling quadratically with circuit length in the worst case.
Furthermore, coherent errors of physical gate operations lead to coherent errors of logical (error-corrected) gate operations such that logical failure scales quadratically with the length of a logical circuit~\cite{Gutirrez2016}.

There exist schemes to mitigate non-Markovian noise by combining open loop control schemes with quantum error correction (QEC).  Examples include dynamical decoupling~\cite{Viola1999,Khodjasteh2005,Biercuk2009,Quiroz2013,Lidar2014,Qi2017,PSL13,Zeng2018}, dynamically corrected gates~\cite{Khodjasteh2009,Khodjasteh2010}, and randomized compiling~\cite{Wallman2016}.
These schemes lower the non-Markovian noise at the expense of additional discrete independent noise and longer operation times.
A lower-cost scheme, applicable when gates are derived from flexible pulse controls, is to compile circuits with hidden inverses that cancel coherent errors without increasing circuit length~\cite{Zhang2022}.  But syndrome extraction itself can convert coherent, physical errors into incoherent, logical errors for sufficiently large codes under certain conditions~\cite{Greenbaum2017, Bravyi2018, Beale2018, Huang2019, Iverson2020, Venn2023, Behrends2025}, calling into question whether and when additional mitigation is necessary.

In this paper, we find:
(1) the average logical failure ``rate'' can increase as a function of the number of error correction rounds if coherent errors are allowed to build constructively;
(2) if the $X$ and $Z$ noise are comparable and one performs passive error correction through virtual Pauli frame updates, then the coherent noise will not build constructively across error correction rounds;
(3) by also initializing the logical qubit in a random Pauli frame, the performance of the logical qubit with correlated noise will essentially be the same as the performance with uncorrelated Pauli noise having the same process fidelity apart from constructive/destructive interference within each round of error correction.  These findings are in contrast to local noise model threshold theorems~\cite{TB05, AGP05, P12} that are quadratically penalized, and they are generally applicable to stabilizer codes without being limited to the asymptotics of large code sizes.  With a ``lazy'' strategy of initializing in a random Pauli frame and performing passive error correction (and assuming comparable $X$ and $Z$ noise), the penalty for having correlated noise is essentially eliminated (with exception to Hamiltonian errors that may add constructively within error correction cycles).

The rest of the paper is as follows: In section~\ref{sec:noise} we describe the noise models we study.  In section~\ref{sec:active} we analyze the failure probability of logical idle in the code capacity model in the context of active error correction.  In section~\ref{sec:passive} we analyze the benefits of initializing in a random codespace and performing ``passive'' error correction using Pauli frame updates.  In section~\ref{sec:realistic} we discuss different time-correlated noise models and caveats
when syndrome extraction is performed using gates that have correlated noise (i.e., a circuit model description).  In section~\ref{sec:code_capacity} 
%and \ref{sec:circuit_model}
we present simulations, corroborating our findings, for code capacity models.%and circuit models respectively.
In section~\ref{sec:conclusion} we present our conclusions.

\section{Syndrome extraction models and noise Types} \label{sec:noise}

We analyze two fundamentally different noise types in this report.  The first is called {\bf Hamiltonian noise} and is described as a Hamiltonian acting on the system.  
\begin{equation}
H_N(t) = \sum_{q}A_q(t) + \sum_{q,r}B_{q,r}(t)
\end{equation}
where $A_q$ is a single qubit random Hamiltonian acting on qubit $q$ and $B_{q,r}$ is a two qubit random Hamiltonian acting on two qubits $q,r$.  We will assume the one and two qubit Hamiltonians are "small" i.e., $\vert \vert \exp(iH) - I \vert \vert_2 \le \delta <<1$, where $H \in \{A_q, B_{q,r} \}$.  Note that the one and two qubit Hamiltonians could be stochastic processes.  Examples of this kind of noise are random small rotation errors in one and two qubit gates and global rotation errors due to shared controls or fields.

A second type of noise we study is the well known {\bf discrete noise }.  This noise is not described by a Hamiltonian but rather as a discrete stochastic process.  This type of noise acts on qubits at discrete times and applies zero-duration large unitaries ($ \vert \vert U - I \vert \vert_2 = \Theta(1)$) to random subsets of qubits.  Note that this is unphysical but nevertheless a popular noise model due to it being efficiently simulatable.  Examples of this kind of noise are Pauli errors on gates, depolarized noise and adversarial stochastic noise studied in \cite{AGP05}.

There are three standard syndrome extraction models (SEM) that model real systems with increasing accuracy.   The first SEM called the {\bf code capacity model} (CCM) assumes syndrome extraction is instantaneous, perfect and does not introduce any errors to the logical qubit.  The second SEM called the {\bf Phenomenological Model} also assumes syndrome extraction is instantaneous.  However each syndrome bit is erroneous and the syndrome errors are an independent Bernoulli process.  The third SEM called the {\bf Circuit Model} (CM) assumes syndrome extraction is performed using one and two qubit gates between encoded qubits and ancillary qubits.  The gates including idles are noisy and can lead to correlated errors among the qubits of the code.

Note that we could combine any of the SEMs with any of the noise types.  Sections \ref{sec:active} and \ref{sec:passive} are discussed in the context of the CCM.  Section \ref{sec:realistic} sketches out a generalization of our arguments to the CM.  We present simulation results for the CCM %and CM
in Sec.~\ref{sec:code_capacity}.
%and Sec.~\ref{sec:circuit_model} respectively.

\section{Active error correction analysis \label{sec:active}}

Consider the evolution of a logical state $\lvert \bar{\psi} \rangle$ encoded in a distance $d$ code in the presence of $X$ noise and error correction in the CCM.  We analyze the evolution of a logical state.  The logical state is subject to noise for one time step and subsequently undergoes syndrome extraction and error correction.  We call the combination of noise for one time step followed by syndrome extraction and error correction a QEC cycle.  The state after a single QEC cycle is given by:
\begin{equation}
\lvert \bar{\psi}_1 \rangle = \left( \lvert \bar{\psi} \rangle + \alpha \bar{X} \lvert \bar{\psi} \rangle \right) / \sqrt{1 + \alpha^2} \label{eq:psiCC}
\end{equation}
where the coefficient $\alpha$ is a random variable whose distribution is a function of the noise and the error correction code.
Consider the coherent evolution of a logical state given by \eq{psiCC} under multiple QEC cycles.  The unnormalized state after $n=2$ applications of Hamiltonian noise and error correction is given by 
\begin{equation}
\lvert \bar{\psi}_2 \rangle = \lvert \bar{\psi} \rangle + \alpha_1 \bar{X} \lvert \bar{\psi} \rangle + \alpha_2 \bar{X} (\lvert \bar{\psi} \rangle + \alpha_1 \bar{X} \lvert \bar{\psi} \rangle )
\end{equation}

For general $n$ the unnormalized state  is given by: 
\begin{equation}
\begin{split}
&\lvert \bar{\psi}_n \rangle =  \left(1 + \sum_{k \ge 1}^n \sum_{1 \leq i_1 < \ldots < i_{2k} \leq n} \alpha_{i_1} \ldots \alpha_{i_{2k}} \right) \lvert \bar{\psi} \rangle \\
&+  \left(\sum_{i=1}^n \alpha_i + \sum_{k \ge 1}^n \sum_{1 \leq i_1 < \ldots < i_{2k+1} \leq n} \alpha_{i_1} \ldots \alpha_{i_{2k+1}} \right)  \bar{X} \lvert \bar{\psi} \rangle \\
&~~~~~~~~\approx \lvert \bar{\psi} \rangle + \sum_{i=1}^n \alpha_i \bar{X} \lvert \bar{\psi} \rangle,
\end{split} \label{eq:stateN}
\end{equation}
where the approximation is the lowest order contribution with respect to $\alpha_i$ factors for each orthogonal term.

Of interest is the average failure probability, $P_{F}(n) = E[1- \vert \langle \bar{\psi_n} \vert \bar{\psi} \rangle \vert ^2]$, where the notation $E[A]$ is the expectation value of random variable $A$.  Then $P_{F}(n)$ for small $n$ is well approximated as:
\begin{equation}
\begin{split}
P_{F}(n) & \approx E\left[\left \vert \sum_{t=1}^{n} \alpha_t \right \vert ^2\right] \label{eq:pfN}\\
& = E\left[\sum_{t=1}^{n} \vert \alpha_t \vert^2 + \sum_{s \neq t } \alpha_s\alpha^*_t\right] 
\end{split}
\end{equation}
The approximation follows from the approximation in \eq{stateN}.  This approximation will fail as $n$ gets large, more precisely when $n \gtrsim 1/E[|\alpha|]$.  We will subsequently show that $E[|\alpha|] \simeq \mathcal{O}(p^{(d+1)/2})$ for local noise where $p$ is the effective physical gate error probability.  So our approximation is valid when $n \ll \mathcal{O}(p^{(d+1)/2})$.  This does not constrain the analysis since $\mathcal{O}(p^{(d+1)/2})$ is the upper bound on the number of encoded gates one is expected to run with a distance $d$ code.

If we assume $\alpha_s$ and $\alpha_t$ are independent for $s \neq t$ then \eq{pfN} simplifies to
\begin{equation}
P_{F}(n) \approx n E[ \vert \alpha \vert ^2] + (n^2-n) \vert E[\alpha] \vert ^2 \label{eq:pf1} 
\end{equation}
If we do not assume the $\{\alpha_t \}$ are independent then we would require a model to derive the cross-correlations ($E[ \alpha_s\alpha^*_t]$) over time.
Note that \eq{pf1} makes no assumption on the underlying noise type and is valid in the CCM (for which the product of the error and the correction is always a logical operator).  $P_{F}(n)$ depends on the noise type only through the mean and standard deviation of $\alpha$.   We will derive these for local and global continuous noise models, as well as a discrete noise model, in the following subsections.

\subsection{Local Noise (LN)} \label{sec:Hln}
Consider a noise Hamiltonian given by $H_{LN} = \sum_{q} \epsilon_q X_q $ where the index $q$ runs over the qubits and $\{\epsilon_q\}$ is an independent, identically, distributed (i.i.d.) sequence with mean $\mu_{\epsilon} $ and variance $\sigma_{\epsilon}^2$.  When a logical state evolves for one time step under this Hamiltonian the state becomes $\lvert \tilde{\psi} \rangle := U_{LN} \lvert \bar{\psi} \rangle$ where 
\begin{equation}
U_{LN} = \otimes_q e^{-i \epsilon_q X_q} \label{eq:Uln} \approx \otimes_q (\idop + \epsilon_q X_q).
\end{equation} 
The approximation is equivalent up to higher orders of $\epsilon$ in each orthogonal operation.
The actual distribution of $\alpha$ to use in \eq{pf1} is dependent on the stabilizer code in addition to the noise model.  We 
start by considering a bit-flip repetition code and then generalize to any code.
Up to higher orders of $\epsilon$ in each orthogonal term (which would be required for proper normalization), the repetition code wavefunction after one step becomes
\begin{equation}
\begin{split}\lvert \tilde{\psi_1} \rangle &\approx \mathcal{O}(1) \lvert \bar{\psi} \rangle + \epsilon_1 \epsilon_2 \cdots \epsilon_d \bar{X} \lvert \bar{\psi} \rangle \\
& + \sum_{i} \epsilon_i X_i  \lvert \bar{\psi} \rangle + \frac{\epsilon_1 \epsilon_2 \cdots \epsilon_d}{\epsilon_i}   X_i \bar{X} \lvert \bar{\psi} \rangle \\
& + \sum_{i,j} \epsilon_i \epsilon_j X_iX_j \lvert \bar{\psi} \rangle + \frac{\epsilon_1 \epsilon_2 \cdots \epsilon_d}{\epsilon_i \epsilon_j}  X_iX_j \bar{X} \lvert \bar{\psi} \rangle \\
 & \phantom{+} \vdots \\
 & + \sum_{i_1,\ldots,i_{\frac{d-1}{2}}} \left( \epsilon_{i_1} \ldots \epsilon_{i_{\frac{d-1}{2}}} X_{i_1} \ldots X_{i_{\frac{d-1}{2}}} \lvert \bar{\psi} \rangle  \right. \\  
 & + \;\;\; \left. \epsilon_{i_{\frac{d+1}{2}}} \ldots \epsilon_{i_d} X_{i_{\frac{d+1}{2}}}..X_{i_d} \bar{X} \lvert \bar{\psi} \rangle \right) \label{eq:alpha_grp}.
\end{split}
\end{equation}

As the Hamming weight of the correction operator increases, the probability of observing the corresponding syndrome decreases and the relative amplitude of the uncorrectable portion of the wavefunction increases.
It follows from \eq{alpha_grp} that the distribution of $\alpha$ for the LN model is, in leading orders of $\epsilon$, given by
\begin{equation}
\label{eq:Rcc}
\alpha_{\rm LN} \approx \left\{ \begin{array}{lll}
& \Pi_{i=1}^d \epsilon_i  \mbox{ with prob. }  \mathcal{O}(1) \\
& \frac{1}{\epsilon_j} \Pi_{i \neq j} \epsilon_i  \mbox{ with prob. }  \epsilon_j^2 \\
& \frac{1}{\epsilon_{i_1}\epsilon_{i_2}} \Pi_{i \neq \{i_1,i_2\} } \epsilon_i  \mbox{ with prob. }  \epsilon_{i_1}^2\epsilon_{i_2}^2 \\
& \vdots  \\
& \frac{\epsilon_{i_{\left\lfloor d /2 \right\rfloor + 1}} \ldots \epsilon_{i_d}}{\epsilon_{i_1} \ldots \epsilon_{i_{\left\lfloor d/2 \right\rfloor}}} \mbox{ with prob. }  (\epsilon_{i_1} \ldots \epsilon_{i_{\left\lfloor d / 2 \right\rfloor}})^2 \\
\end{array}
\right.
\end{equation}
noting that the probabilistic contribution for each syndrome is determined by the normalized correctable/uncorrectable pair of terms.  
The various indices take values in $\{1, \ldots, d \}$.  
\eq{Rcc} implies 
\begin{align}
E[\vert \alpha_{LN} \vert^2] &\approx a_d E[(\epsilon^2)]^{\frac{d+1}{2}} = a_d (\mu_{\epsilon}^2 + \sigma_{\epsilon}^2)^{\frac{d+1}{2}} \label{eq:valpha_ln} \\
\vert E[\alpha_{LN}] \vert^2 &\approx b_d \mu_{\epsilon}^{2d} \label{eq:malpha_ln} 
\end{align}
where $a_d$ and $b_d$ are real numbers that depend upon the code distance and, when we generalize this beyond the repetition code, upon the code itself.
The former equation is dominated by the last case of \eq{Rcc} and 
$a_d = 
\left(
\begin{array}{c}d\\
\lfloor d / 2 \rfloor
\end{array}
\right)$ for the repetition code.
The latter equation has equal order contributions from each case of \eq{Rcc}
and $b_d = 
\sum_{k=0}^{\lfloor d / 2 \rfloor} \left(
\begin{array}{c}d\\
k
\end{array}
\right)$ for the repetition code.

Proper restrictions on qubit indices can generalize Eq.~(\ref{eq:Rcc}) for any code so that each case corresponds to a set of qubit errors and a complement that combine to a $d$-weight logical operator.
While higher-weight logical operators also make contributions to $\alpha$, the lowest order contributions with respect to $\epsilon$ are covered by this generalization of Eq.~(\ref{eq:Rcc}).
This allows us to predict the qualitative behavior of the CCM failure probability in the low error regime for any code of distance $d$ from Eqs.~(\ref{eq:pf1}),  (\ref{eq:valpha_ln}), and (\ref{eq:malpha_ln}) to obtain:
\begin{equation}
P^{LN}_F(n) \approx n a_d (\mu_{\epsilon}^2 + \sigma_{\epsilon}^2)^{\frac{d+1}{2}} + (n^2-n) b_d \mu_{\epsilon}^{2d} \label{eq:pfniid},
\end{equation}
where $a_d$ is the number of lowest-weight malignant sets of the code and $b_d$ is the number of all 
proper subsets of lowest-weight malignant sets.  This is consistent with our findings that were specific to the repetition code.

Two special cases of the above Hamiltonian are: 
\begin{itemize}
\item $\mu_{\epsilon}=0, \sigma_{\epsilon}>0$ which is equivalent to the discrete noise case.  Here $P_F(n)$ reduces to 
\begin{equation} 
n a_d (\sigma_{\epsilon}^2)^{\frac{d+1}{2}} \label{eq:pfnmu0}
\end{equation}
\item $\mu_{\epsilon}>0, \sigma_{\epsilon}=0$ which is equivalent to a consistent, global over-rotation noise.  Here $P_F(n)$ reduces to 
\begin{equation}  
n a_d (\mu_{\epsilon}^2)^{\frac{d+1}{2}} + (n^2-n) b_d \mu_{\epsilon}^{2d} \label{eq:pfnsigma0}
\end{equation}
\end{itemize}

%%%AG%%%

\subsection{Global Noise (GN)} \label{sec:Hgn}

Consider a noise Hamiltonian given by $H_{GN} = \sum_{q} \epsilon X_q $ where the index $q$ runs over the qubits and $ \epsilon $ is a random variable with mean $\mu_{\epsilon} $ and variance $\sigma_{\epsilon}^2$.  When a logical state evolves for one time step under this Hamiltonian the state becomes $\lvert \tilde{\psi} \rangle := U_{GN} \lvert \bar{\psi} \rangle$ where 
\begin{equation}
U_{GN} = \otimes_q e^{-i \epsilon X_q}
\end{equation} 
The actual distribution of $\alpha$ is dependent on the stabilizer code in addition to the noise model and its derivation is very simlar to that of the local noise $H_{LN}$ model.  We will reuse arguments we presented in that analysis for generalization and focus on the repetition code for simplicity.  We rewrite the state $U_{GN} \lvert \bar{\psi} \rangle$ by grouping terms with a common syndrome for a distance $d$ bit flip repetition code.  Up to higher orders of $\epsilon$ in each orthogonal term, we have
\begin{equation}
\begin{split}
\lvert \tilde{\psi} \rangle &\approx \mathcal{O}(1) \lvert \bar{\psi} \rangle + \epsilon^d  \bar{X} \lvert \bar{\psi} \rangle \\
& + \sum_{j} \epsilon X_j  \lvert \bar{\psi} \rangle + \epsilon^{d-1}  X_j \bar{X} \lvert \bar{\psi} \rangle \\
& + \sum_{i,j} \epsilon^2 X_iX_j \lvert \bar{\psi} \rangle + \epsilon^{d-2} X_iX_j \bar{X} \lvert \bar{\psi} \rangle \\
 & \phantom{+} \vdots \\
 & + \sum_{i_1,\ldots,i_{\frac{d-1}{2}}} \left( \epsilon^{\frac{d-1}{2}} X_{i_1} \ldots X_{i_{\frac{d-1}{2}}} \lvert \bar{\psi} \rangle  \right. \\  
 & + \;\;\; \left. \epsilon^{\frac{d+1}{2}} X_{i_{\frac{d+1}{2}}}..X_{i_d} \bar{X} \lvert \bar{\psi} \rangle \right) \label{eq:alpha_grp_gn}
\end{split}
\end{equation}
It follows from \eq{alpha_grp_gn} the distribution of $\alpha$ for a distance $d$ repetition code under Global Noise Hamiltonian, in leading orders of $\epsilon$, is given by
\begin{equation}
\alpha_{R, GN} = \left\{ \begin{array}{lll}
& \epsilon^{d}  \mbox{ with prob. }  \mathcal{O}(1) \label{eq:Rcc_gn}\\
& \epsilon^{d-2} \mbox{ with prob. } \epsilon^2 \\
& \epsilon^{d-4} \mbox{ with prob. } \epsilon^4 \\
& \vdots  \\
& \epsilon^1 \mbox{ with prob. } \epsilon^{d-1} 
\end{array}
\right.
\end{equation}
\eq{Rcc_gn} implies 
\begin{align}
E[\alpha_{GN}^2] &= a_d E[\epsilon^{d+1}] \label{eq:valpha_gn} \\
\vert E[\alpha_{GN}] \vert^2 &= b_d \vert E[\epsilon^{d}] \vert^2 \label{eq:malpha_gn}
\end{align}
where $a_d$ and $b_d$ are the same real numbers as in \secref{Hln}.
Hence the failure probability is given by
\begin{equation}
P^{GN}_F(n) \approx n a_d E[\epsilon^{d+1}] + (n^2-n) b_d \vert E[\epsilon^d] \vert^2 \label{eq:pf_gn}
\end{equation}
Note that \eq{pf_gn} and \eq{pfnsigma0} are identical when $\sigma_{\epsilon} = 0$.
%%%AG%%%
\subsection{Discrete noise} \label{sec:discrete}
This is the well studied model where the noise is described as a stochastic process occurring at discrete time steps on each qubit.  We consider the case where each qubit independently of the other qubits gets an $X$ error with probability $p$.  
This is followed by perfect syndrome extraction and error correction.  At the end of a QEC cycle we would either have the original state or it would be rotated by $\bar{X}$.  In the case of a distance $d$ code the probability of this event is $\mathcal{O}(p^\frac{d+1}{2})$.  More specifically, this probability, to lowest order, is $a_d p^\frac{d+1}{2}$ where $a_d$ is the number of minimum-weight malignant sets, consistent with its definition in \secref{Hln} and \secref{Hgn}.  Hence:
\begin{equation}
P_F^D(n) \approx n a_d p^\frac{d+1}{2} \label{eq:pfdn}.
\end{equation}
This is a good approximation when $n \ll \mathcal{O}(p^{(d+1)/2})$ such that there is a low probability of multiple failures.
Note that \eq{pfdn} and \eq{pfnmu0} are identical if we take $p = \sigma_{\epsilon}^2$ which is the average Born-rule probability of a flip after applying $e^{-i \epsilon X}$ to a polarized qubit.

\section{Benefits of passive error correction \label{sec:passive}}
\subsection{Random initial codespace}
\label{sec:randinit}
A $[[n,k,d]]$ stabilizer code is a subspace, called the codespace, of an $n$ qubit Hilbert space.  The codespace is customarily chosen to be the joint '+1' eigenspace, called the syndrome-zero subspace, of stabilizers.  Pauli errors that can be detected move the encoded state to a non-zero syndrome subspace.  However, in the presence of coherent noise it is beneficial to let the codespace actually be a non-zero syndrome subspace.  Consider the case of a code subjected to continuous, uniform $X$-noise.  When the code is in the syndrome-zero subspace, the noise rotations can add constructively in the logical error space.  In the non-zero $X$-syndrome subspace, there can be destructive interference to mitigate the noise effect.

To be more concrete, consider two uncorrectable Pauli-$X$ errors with the same syndrome in the expansion of the coherent $X$-noise operator $\exp(iH_{\mbox{noise}}) = \ldots a_1 E_1^X + a_2E_2^X \ldots$.  If $E_1^X E^X_2$ is a stabilizer then the erroneous post-correction state given this syndrome will have the term $(a_1 + a_2) \bar{X} | \bar{\psi} \rangle$ from these contributions.  This error contribution can be impacted by the codespace to the extent that $|a_1 + a_2|$ differs from $|a_1 - a_2|$.  Initializing a logical qubit into a random codespace can be straightforward.  For example, in a qLDPC code one may simply initialize physical qubits at random in $\{\ket{0}, \ket{1} \}$ or $\{\ket{+}, \ket{-} \}$, depending on $\ket{\bar{0}}$ or $\ket{\bar{+}}$ preparation, and start applying rounds of syndrome extraction to these physical qubits and the state will collapse into a random codespace.  The syndrome measurements as well as the initial physical states indicate which one.
Our simulations on the rotated surface code in Sec.~\ref{sec:code_capacity}
%and \ref{sec:circuit_model}
show that there is a substantial benefit from starting in a random codespace for combating Hamiltonian noise.

\subsection{Random codespace walk}
\label{sec:randwalk}
There is an additional benefit if one also allows the codespace to wander in a random walk fashion by tracking the Pauli frame passively rather than actively correcting errors.
Consider the continuous noise model with $X$ noise and a distance $d$ stabilizer code.  Recall that the evolution after $N$ QEC cycles is approximated as
$\lvert \bar{\psi}_n \rangle \approx \lvert \bar{\psi} \rangle + \sum_i \alpha_i \bar{X} \lvert \bar{\psi} \rangle$ [\eq{stateN}].
Suppose, in addition, we have a discrete time $Z$ error process on the qubits.  This has the effect of suppressing the $\bar{X}$ amplitude on the encoded space.  To see this, suppose qubits $i$ and $j$ had $Z$ errors at times $t_1$ and $t_2>t_1$ with $Z_i$ anticommuting with $\bar{X}$.  Then the effective operator acting on the initial $\lvert \bar{\psi} \rangle$ is
\begin{equation}
Z_j (I + s_2 \bar{X}) Z_i (I+s_1 \bar{X}) = (1-s_1s_2)Z_iZ_j + (s_1-s_2)Z_iZ_j \bar{X} \label{eq:Zdd}
\end{equation}
where $s_1$ and $s_2$ are random variables that correspond to the coherent error built in the $\bar{X}$ in time $[0,t_1]$ and $[t_1, t_2]$ respectively.  
Assuming that $Z_i$ and $Z_j$ can be identified and corrected, they are effectively identity operations (post-correction).
One should not actively correct these as they are detected, but rather do the classical bookkeeping of Pauli frame updates.  In the corrected Pauli frame, the total operator in \eq{Zdd} becomes
\begin{equation}
(1-s_1s_2)I + (s_1-s_2) \bar{X}
\end{equation}

More generally, if we had stochastic $Z$ noise with probability $p$ in addition to continuous $X$ noise then the coherent noise does a sort of a one dimensional random walk on the real line with a step size distributed as $\alpha$ and a direction reversal after $R$ steps where $R$ is distributed geometrically with mean $1/p$.  Consequently,
\begin{equation}
E[R] = \frac{1}{p},~E[R^2] - E[R]^2 = \frac{1-p}{p^2}. \label{eq:Rstats}
\end{equation}
If we look at $2k$ direction reversals of this walk then the total displacement is
\begin{equation}
\Gamma_{2k} = S_1 -S_2 + \ldots + S_{2k-1} - S_{2k} \label{eq:gamma}
\end{equation}  
where 
\begin{equation}
S_i = \sum_{j=1}^{R_i} \alpha_{i,j} \label{eq:randsum}
\end{equation}
and $\{R_i\}$ are i.i.d. geometric random variables governed by Eqs.~(\ref{eq:Rstats}).
Let $Q_F(2k)$ be the probability of failure after $2k$ steps so that $Q_F(2k) \approx E[|\Gamma_{2k}|^2]$, applying the same approximation as that of \eq{pfN} and assuming that all of the $Z$ error processes are correctable.
\eq{gamma} implies $\Gamma_{2k}$ is a sum of $k$ zero mean independent random variables and we have
\begin{equation}
Q_F(2k) \approx E[|\Gamma_{2k}|^2] = kE[\vert S_1 - S_2 \vert ^2] = 2k E[|S_1|^2]. \label{eq:mgamma}
\end{equation}
We will compute $E[|S_1|^2]$ using \eq{randsum}:
\begin{equation}
\begin{split}
E[|S_1|^2] &= E_R \left[ R E[|\alpha|^2] + (R^2-R) E^2[|\alpha|] \right] \\
&= E[R] E[|\alpha|^2] + (E[R^2] - E[R])  E^2[|\alpha|]. \label{eq:meanS}
\end{split}
\end{equation}
Employing Eqns.~(\ref{eq:Rstats}), (\ref{eq:mgamma}) and (\ref{eq:meanS}), we then have
\begin{equation}
Q_F(2k) \approx 2k E[R] (E[|\alpha|^2] + \frac{2 - 2 p}{p} E^2[|\alpha|]). \label{eq:q2k}
\end{equation}

In terms of the average error as a function of the number of QEC rounds, the failure probability
when there is discrete noise with probability $p$ that anti-commutes with Hamiltonian noise is
\begin{equation}
P^{p}_F \approx E[Q_F(n/E[R])] \approx n (E[|\alpha|^2] + \frac{2 - 2 p}{p} E^2[|\alpha|]) \label{eq:pfp}
\end{equation}
since $E[2k\mbox{ steps}]=2kE[R]$ QEC cycles.
Comparing this with the active error correction scenario of \eq{pf1}, the first term is equivalent but the second term loses its quadratic dependence on $n$.  This results because the random walk effect suppresses a build-up of the biased coherent error.

We have already computed $E[|\alpha|^2]$ and $E^2[|\alpha|]$ for the $H_{LN}$ and $H_{GN}$ cases in \secref{Hln} and \secref{Hgn} respectively.  
\subsection{Local noise}
From Eqs.~(\ref{eq:pfp}), (\ref{eq:valpha_ln}), and (\ref{eq:malpha_ln}) in the $H_{LN}$ case we have
\begin{equation}
P^{pLN}_F \approx n \left( a_d (\mu_{\epsilon}^2 + \sigma_{\epsilon}^2)^{\frac{d+1}{2}} + \frac{2-2p}{p} b_d \mu_{\epsilon}^{2d} \right). \label{eq:pf_pLN}
\end{equation}
 If the discrete time $Z$ noise has a comparable strength to the bias of the continuous time $X$ noise then $p \sim \mu_{\epsilon}^2$ and the failure probability reduces to
 \begin{equation}
 P^{pLN}_F(n) \approx n \left( a_d (\mu_{\epsilon}^2 + \sigma_{\epsilon}^2)^{\frac{d+1}{2}} + b_d\mu_{\epsilon}^{2d-2} \right) \label{eq:pf_pLN2}.
\end{equation}
Not only is the second term linear in $n$, unlike the active error correction scenario of \eq{pf1}, but it is a higher order term when $d > 3$.  Therefore, this model becomes equivalent, in the lowest order, to the unbiased local noise model of \eq{pfnmu0} with $\sigma' = \sqrt{\mu^2 + \sigma^2}$ or a discrete noise model of \eq{pfdn} with $p = \mu^2 + \sigma^2$. 

\subsection{Global Noise}
\label{sec:passive_gn}
From Eqs.~(\ref{eq:pfp}), (\ref{eq:malpha_gn}), (\ref{eq:valpha_gn}),  in the $H_{GN}$ case we have
\begin{equation}
P^{pGN}_F(n) \approx n \left( a_b E[\epsilon^{d+1}] + \frac{2 - 2p}{p} b_d \vert E[{\epsilon^d}] \vert^2 \right) \label{eq:pf_pGN}.
\end{equation}
The first term is unchanged compared with \eq{pf_gn} but the second term no longer scales quadratically with $n$.

If we assume that $\epsilon$ is normally distributed with $\mu_{\epsilon} = 0$, the odd moments are zero and the even moments are known to be $E[\epsilon^{2k}] = (2k-1)!!~\sigma_{\epsilon}^{2k}$ where $!!$ denotes the double factorial defined such that $(2k-1)!! = (2k-1)(2k-3) \ldots 1$.  Then we have
\begin{equation}
P^{pGN}_F(n) \approx d!!~n a_b \sigma_{\epsilon}^{d+1} \label{eq:pf_pGN2},
\end{equation}
assuming $d$ to be odd.  Here we see a penalty of a $d!!$ factor from the variation and spatial correlation of $\epsilon$.
However, in a noise model that limits the possible values of $\epsilon$ to some $\epsilon_0$, we would have $P^{pGN}_F(n) < n_a \epsilon_0^{d+1}$.

\subsection{Discrete versus Hamiltonian noise}
\label{sec:continuous_orthogonal}

So far we have analyzed the noise suppression of discrete $Z$-noise on Hamiltonian $X$-noise.  Suppose instead we had Hamiltonian $Z$(or $Y$)-noise in addition to Hamiltonian $X$-noise.  The act of measuring $X$-type stabilizers converts the continuous $Z$-noise to a discrete $Z$ operator on a single qubit with a frequency of $O(\epsilon^{-2})$.  So stabilizer measurements convert continuous $Z$-noise with strength $\epsilon$ per time step into discrete-$Z$ noise with probability of $\epsilon^2$.  So, if we use passive error correction and the quantum system has Hamiltonian $X$- and $Z$(or $Y$)-noise then the coherent build-up of $X$-noise is suppressed due to $Z$(or $Y$)--noise and vice-versa and we expect $P_F(n)$ to scale linearly with $n$.

Combining an initial random codespace and a random codespace walk in a $d>3$ code, we find that a Hamiltonian noise model is as benign as a discrete noise model with fault probabilities set to match the Hamiltonian noise model after one round.  This is observed in simulations that are presented in \secref{code_capacity}.

\section{Realistic noise models \label{sec:realistic}}
\subsection{Time correlated noise}

So far we have analyzed the case where the errors are independent across time.  Consider the local noise Hamiltonian $H_{LN}(n)  = \sum_{q} \epsilon_q(n) X_q $ where $n$ represents a discrete time index.  We will now assume that there are time correlations amongst $\{ \epsilon_q(n)\}$.  We will consider a simple model of correlation: $$\epsilon_q(n+1) = \sqrt{\beta} \epsilon_q(n) + \sqrt{1-\beta} \delta_q(n+1)$$  
\begin{itemize}
\item $\beta$ is a parameter in $[0,1]$ that sets the strength of the time correlations 
\item $\{ \delta_q(n) \}_{q,n}$ are i.i.d random variables
\end{itemize}
We note that this model is a special case of the local non-Markovian noise model considered in~\cite{AGP05}.  In that article, there was no assumption made on the structure of the temporal correlations.

The following sub-sections analyze the case of perfect temporal correlation ($\beta=1$) under two different extremes with respect to spatial correlations.  Our local noise model assumes that each qubit noise is independent.  Our global noise model assumes that the same noise acts on all qubits (perfectly correlated in space and time.  These extremes are convenient to study analytically and cover the possible worst-case scenarios.  We have confirmed this in our simulations presented in Sec.~\ref{sec:code_capacity}.

\subsubsection{Time correlated local noise}
\label{sec:tcln}

In the case where there is perfect temporal correlation ($\beta = 1$) but perfectly uncorrelated in space, we can derive an approximation to the failure probability $P^{TC-LN}_F(n)$ in the time-correlated local noise model by following the analysis we did in \secref{Hln}.  We will use \eq{pfN} to approximate $P^{TC-LN}_F(n)$.  We need to compute $E[|\alpha_t|^2]$ and $E[|\alpha_t \alpha_s|]$.  We already have $E[|\alpha_t|^2] = a_d (\mu^2_{\epsilon} + \sigma^2_{\epsilon})^\frac{d+1}{2}$ from \eq{valpha_ln}.  We can compute $E[|\alpha_t \alpha_s|]$ using \eq{Rcc}.  By averaging over pairs of syndrome possibilities for time steps $s$ and $t$ and setting $\epsilon_q(s)= \epsilon_q(t)$ we obtain $E[|\alpha_t \alpha_s|] = b_d E[(\epsilon^2)]^d = b_d (\mu^2_{\epsilon} + \sigma^2_{\epsilon})^d$.  Using these equalities we then have:
\begin{align}
P^{TC-LN}_F(n) & \approx n E[\vert \alpha_1 \vert^2] + (n^2-n)E[\vert \alpha_1\alpha_2 \vert]  \nonumber \\
& \approx n a_d (\mu^2_{\epsilon} + \sigma^2_{\epsilon})^\frac{d+1}{2} \nonumber \\
& + (n^2-n) b_d (\mu^2_{\epsilon} + \sigma^2_{\epsilon})^d \label{eq:pfn_tcln}
\end{align}

If we introduce orthogonal, discrete $Z$ noise with probability $p$ in the passive error correction setting, we can use the analogue of \eq{pfp} with time correlation to obtain
\begin{align}
P^{pTC-LN}_F(n) & \approx n \left[ E[ \vert \alpha_1 \vert^2] + \frac{2 - 2p}{p} E[\vert \alpha_1\alpha_2 \vert] \right]  \nonumber \\
& \approx n \left[ a_d (\mu^2_{\epsilon} + \sigma^2_{\epsilon})^\frac{d+1}{2}) \right. \nonumber \\
& \left. + \frac{2 - 2 p}{p} b_d  (\mu^2_{\epsilon} + \sigma^2_{\epsilon})^d \right] \label{eq:pfn_ptcln}.
\end{align}
If we assume the $Z$ noise has a comparable strength as the $X$ noise, then $p \sim \mu^2_{\epsilon} + \sigma^2_{\epsilon}$ and the failure probability reduces to
\begin{equation}
P^{pTC-LN}_F(n) \approx n  \left[a_d ( \mu^2_{\epsilon} + \sigma^2_{\epsilon})^\frac{d+1}{2} + b_d (\mu^2_{\epsilon} + \sigma^2_{\epsilon})^{d-1} \right] \label{eq:pfn_ptcln2}.
\end{equation}
We again find that the second term is a higher order correction when $d>3$ and the first term is equivalent to the discrete noise model of \eq{pfdn} with $p = \mu^2_{\epsilon} + \sigma^2_{\epsilon}$.  As argued before, the orthogonal noise need not be discrete.  With continuous noise that is comparable in two orthogonal directions, each has the effect of suppressing the coherent build-up of noise in the other.

\subsubsection{Time correlated global noise}
\label{sec:tcgn}

In the extreme limit that the noise is perfectly correlated over time and space over a logical qubit, the failure probability ($P^{TC}_F(n)$) analysis is essentially the same as the global noise from \secref{Hgn}.  We need to compute $E[|\alpha_t|^2]$ and $E[|\alpha_t \alpha_s|]$.  We already have $E[|\alpha_t|^2] = a_d E[\epsilon^{d+1}]$ from \eq{valpha_gn}.  We can compute $E[|\alpha_t \alpha_s|]$ using \eq{alpha_grp_gn}.  By averaging over pairs of syndrome possibilities for time steps $s$ and $t$ and setting $\epsilon(s)= \epsilon(t) = \epsilon$ we obtain $E[|\alpha_t \alpha_s|] = b_d E[\epsilon^{2d}]$.  Using these equalities we then have:
\begin{align}
P^{TC-GN}_F(n) & \approx n E[\alpha_1^2] + (n^2-n)E[\alpha_1\alpha_2]  \nonumber \\
& = n a_d E[\epsilon^{d+1}]) + (n^2-n) b_d E[\epsilon^{2d}] \label{eq:pfn_tcgn}.
\end{align}
Further assuming the $\epsilon$ is normally distributed with $\mu_{\epsilon} = 0$ as we did to obtain \eq{pf_pGN2}, we have
\begin{equation}
P^{TC-GN}_F(n) \approx n d!!~a_d \sigma_{\epsilon}^{d+1}  + (n^2-n) (2d - 1)!!~b_d \sigma_{\epsilon}^{2d} \label{eq:pfn_tcgn_normal}
\end{equation}

If we introduce discrete $Z$ noise with probability $p$ in the passive error correction setting, we can use the analogue of \eq{pfp} with time correlation to obtain
\begin{align}
P^{pTC-GN}_F(n) & \approx n \left[ E[\alpha_1^2] + \frac{2 - 2p}{p} E[\alpha_1\alpha_2] \right]  \nonumber \\
& \approx n \left[ a_d E[\epsilon^{d+1}]) + \frac{2 - 2p}{p} b_d E[\epsilon^{2d}] \right]. \label{eq:pfn_ptcgn}
\end{align}
Assuming $\epsilon$ is normally distributed with $\mu_{\epsilon} = 0$
and also assuming that the orthogonal noise strength are comparable with $p = \mathcal{O}(E[\epsilon^2])$, we have
\begin{equation}
P^{pTC-GN}_F(n) \approx n \left[ d!!~a_d \sigma_{\epsilon}^{d+1} + (2d-1)!!~ b_d \sigma_{\epsilon}^{2d-2} \right]. \label{eq:pfn_ptcgn2}
\end{equation}
Once again, the second term is a higher order correction when $d>3$ (though the pre-factor could be substantial).  
However, we see the same $d!!$ factor penalty from the variation and spatial correlation of $\epsilon$ as in \eq{pf_pGN2}.
Again, this penalty is not present if there is a narrower distribution of $\epsilon$ such as having a hard cutoff.

\subsection{Circuit model (CM)}

We have thus far analyzed the CCM where syndrome extraction is a noise-free one step process.  
In practice syndrome extraction will be a noisy operation composed of many noisy one and two qubit gates that can spread errors.  If the syndrome extraction circuit is fault tolerant, the code will achieve error suppression even with error spreading as seen in the context of the Surface Code and qLDPC codes~\cite{RHG2006, MC2025}.  Correlations of the noisy operations, however, may alter the logical qubit performance.  In principle, this could be analyzed in a perturbative fashion in a Pauli channel representation as we have done in the CCM.  Basically, consider the probabilistic dynamical map of the error correction cycle, including the measurements, and expand in orders of error contributions (e.g., $\epsilon$) in a Pauli error representation.  That is, for each syndrome input there will be a probability distribution of syndrome outputs each with its own linear combination (unitary mapping) of Pauli errors, analogous to the distribution of outcomes expressed in \eq{Rcc} but generalized in this more sophisticated context.  One caveat is that $n$ fault sources can contribute to terms with more than $n$ errors due to error spreading.  For example, there may be $\mathcal{O}(\epsilon)$ terms with multiple Pauli error factors.  Investigating the complexities of the CM in detail is out of scope for this paper.  However, the following are some general comments about how our CCM conclusions generalize to the CM.

Correlations within each QEC round may add constructively (or destructively) and alter the distribution of syndrome outcomes relative to an error model that has uncorrelated noise with the same process fidelity.  As in the CCM, these effects are mitigated by starting in a random Pauli frame.  However, depending upon the length of the syndrome extraction circuit and the nature of the correlations, errors may add constructively in a way that this strategy cannot fully suppress.  Across rounds, the build up of coherent errors will be suppressed using passive error correction using similar arguments that we applied to the CCM.  The random walk nature of passive error correction will regularly flip the sign of error contributions.
Therefore, although noise contributions may build constructively within a QEC round, passive error correction is expected to suppress the quadratic term in $P_F(n)$.  

An equally important question is whether $\delta^2$ is the correct proxy for physical gate error probability in the circuit model if the noise strength (the matrix norm of the Hamiltonian) per gate is bounded by $\delta$.   Since the total number of gates seen by a qubit before a stabilizer measurement is bounded (by some constant $k$) we can upper bound the failure probability assuming a local noise Hamiltonian in the CCM with $\mu_{\epsilon} = k\delta$.  It follows from \eq{pf_pLN2} that, assuming we have both continuous $X$-noise and $Z$-noise and we perform passive error correction,
\begin{equation}
P_F(n) \approx n \mathcal{O}(\delta^{d+1}).
\end{equation}
Comparing the right hand side to the logical failure probability of the discrete noise in \eq{pfdn} we see that $\delta^2$ is indeed a proxy for the physical gate failure probability.

\section{Code capacity model Results \label{sec:code_capacity}}
We have performed vector state simulations of logical idle in the CCM.  We estimate the average success/failure probability as a function of the number of error correction cycles through random sampling of the noise and intermediate measurement projections.  Specifically, we compute statistical instances of $\bar{\psi}_n$ by applying $n$ rounds of QEC to an initially ideal $\bar{\psi}$ using probabilistic noise and quantum measurements.  Then the logical failure after $n$ rounds,
\begin{equation}
P_F(n) = E[\vert \langle \bar{\psi}_n \vert \bar{\psi} \rangle \vert^2],
\end{equation}
is estimated by averaging over the statistical instances. 
The simulations reported in the section confirm our analytical results and resolve any questions about higher order corrections.

\begin{figure}
\includegraphics[height=2.5cm]{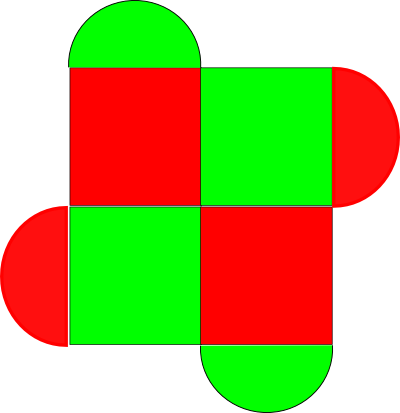}
\caption{A distance three medial surface code.  The qubits are on the vertices and the plaquettes represent stabilizers.  Each of the color represents an $X$-type and a $Z$-type stabilizer.  The particular choice is arbitrary.\label{fig:kbc3}
}
\end{figure}

We study the medial surface code~[\fig{kbc3}] and the bit flip repetition code.  In the former code, the ``success probability'' is defined as the ability to maintain a Bell state between the logical qubit and an ideal ancillary qubit, thereby demonstrating the ability to preserve an arbitrary quantum state.  All four Bell states are equivalent for this purpose.  In the latter code, the ``success probability'' is defined as the ability to maintain an arbitrary initial ``classical'' state.  All ``classical'' states are equivalent for this purpose.  Although the repetition code alone is not useful for protecting quantum information (it only protects against bit flips), the phenomena that we describe is well represented in this model.  We therefore use the repetition code because it is more efficient to simulate than full quantum codes and allows us to go beyond $d=3$ without great computational effort.  Simulations of the surface code in the code capacity and circuit models provide a more complete assessment without altering the conclusions.

\begin{figure}
\includegraphics[height=6cm]{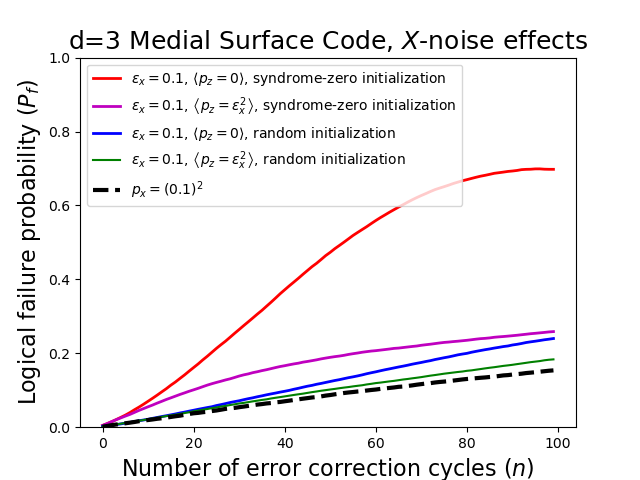}
\caption{For a $d=3$ medial surface code, error correction performance due to continuous/discrete $X$ noise with/without codespace randomization and with/without Pauli-$Z$ flips to induce a random walk in the codespace.  Where present, the Pauli-$Z$ flips are tracked and perfectly corrected.  This is what the angle brackets in the legend are meant to indicate.  The sample uncertainty for each curve is comparable to its thickness.\label{fig:msc3}
}
\end{figure}

In \fig{msc3} we simulate the medial surface code in the CCM with different combinations of global continuous noise with a constant $\epsilon$  (equivalent to local noise with $\sigma_{\epsilon} = 0$ and $\mu_{\epsilon} = \epsilon$) and discrete Pauli flips of probability $p$.  Subscripts of $x$ and $z$ in the legend denote $X$-type and $Z$-type noise.  The plotted failure probabilities only account for the effects from the $X$ noise.  Where present, the discrete Pauli-$Z$ flips are tracked and perfectly corrected at the end so that we can understand its effect purely on the Hamiltonian $X$ noise, testing the random walk described in Sec.~\ref{sec:randwalk} that dampens the coherent build-up of error probability due to the Hamiltonian noise.  These are introduced at probability $p=\epsilon^2$ which is the same probability as an individual spin flip after one application of the Hamiltonian noise with strength $\epsilon$.  That is, these continuous and discrete noise contributions are equivalent apart from coherent noise effects introduced in the Hamiltonian noise component.  Note that the $X$ and $Z$ labels are symmetric for this consideration (the results would be the same if we had continuous $Z$ noise and discrete $X$ noise).  The simulations with nonzero $p_z$ in \fig{msc3} mimic the case of having both $X$ and $Z$ type Hamiltonian noise of comparable strength but isolate the effects of each one independently.  

The worst case scenario in \fig{msc3} is when the logical qubit is initialized in the syndrome-zero code space and there are no Pauli-$Z$ flips ($p_z=0$).  This case exhibits a prominent quadratic scaling of the failure probability as a function of the number of cycles $n$.  It has the worst case initial code space (the syndrome-zero subspace) and has no $Z$ flips to induce a random walk effect.  Introducing Pauli-$Z$ flips that are actively corrected is an equivalent model (by correcting flips, you lose the benefits).  Introducing tracked, discrete Pauli-$Z$ flips that are corrected only at the end, we observe benefits of a random walk effect.
Initializing into the syndrome-zero code space with $p_z = \epsilon_x^2$, we see the same initial steep slope of $P_f$ vs $n$ but a downward curvature due to the benefits of a random walk in the codespace.  There is better initial performance when we start in a random codespace.  However, without introducing Pauli-$Z$ flips, we see a quadratic upward curve in $P_f$ vs $n$.  Finally, when we use a random initial codespace and have the Pauli-$Z$ flips, we see the best performance without much upward curvature.  For comparison, we also show the case when the $X$ noise is discrete.  The difference in these final best cases shows a residual loss of performance due to continuous noise that is expected at $d=3$ according to \eq{pf_pLN2} with $\sigma_{\epsilon} = 0$.  At higher distances, the continuous noise penalty will be suppressed even further.  This comparison is also an indication that the perturbative approximation is appropriate even at a relatively high value of $\epsilon = 0.1$.

\begin{figure}
\includegraphics[height=6cm]{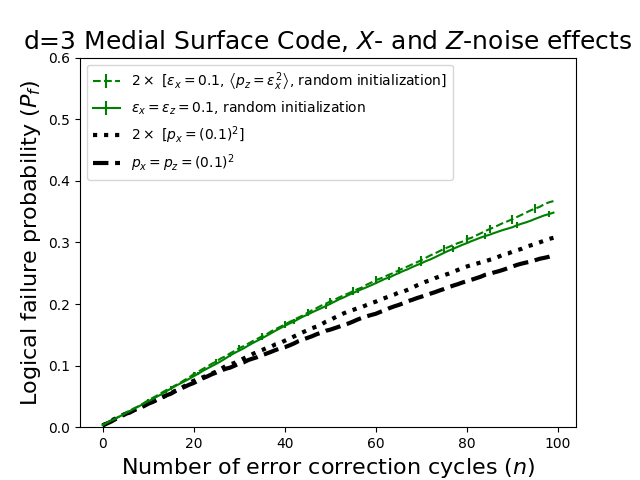}
\caption{For a $d=3$ medial surface code, error correction performance due to continuous/discrete $X$ and $Z$ noise with/without codespace randomization.  The $X$-noise only analogs from \fig{msc3} are re-plotted but multiplied by two for comparison.  The error bars indicate sample uncertainty estimates (where it is significant relative to the thickness of the respective curve). \label{fig:msc3b}
}
\end{figure}

We look at the combined effects of $X$ and $Z$ noise in \fig{msc3b}.  Here, we compare the case of continuous $X$ and $Z$ noise, $\epsilon_x = \epsilon_z = 0.1$, with random initialization to the case of discrete $X$ and $Z$ noise, $p_x = p_z = (0.1)^2$, all with the same lowest order strength.
Also for comparison, we re-plot the equivalent cases with only $X$-effects from ~\fig{msc3} but multiplied by two for a naive doubling of the error from the two noise contributions ($X$ and $Z$).  This doubling of the single channel effect provides a bound of the observed failure probability for the two channels.  The performance when treating the combined noise holistically is slightly better than doubling the single channel because of double-counted failures (failure in both channels represents a single failure instance, not two).  The fact that these are comparable justifies studying the noise effects independently, reinforcing our conclusions from \fig{msc3} as being applicable to a more realistic noise model.  At smaller noise strengths, the effects from the two channels are even more independent (double-counted instances are more rare).

\begin{figure}
\includegraphics[height=6cm]{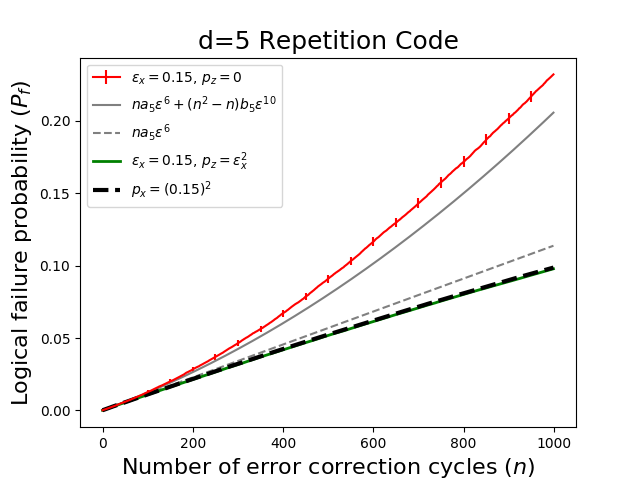}
\caption{For a $d=5$ repetition code, error correction performance due to continuous/discrete $X$ noise with/without Pauli-$Z$ flips.  Also, for comparison, leading order analytical results of \eq{pfnsigma0} (solid gray) and \eq{pf_pLN2} (dashed gray) with $\mu_{\epsilon} = \epsilon$, $\sigma_{\epsilon} = 0$  and using $a_5 = {5 \choose 3} = 10$, $b_5 =  {5 \choose 3} +  {5 \choose 4} +  {5 \choose 5} = 16$ for the repetition code. 
Pauli-$Z$ flips have no effect on the fidelity of states in this non-quantum code, so these do not need to be corrected.  They do, however, flip the relative signs of $\left\lvert\bar{\psi}\right\rangle$ and $\bar{X} \left\lvert\bar{\psi}\right\rangle$, providing a random walk effect to suppress the penalty of continuous noise.
Error bars indicate sample uncertainty estimates (where it is applicable and significant relative to the thickness of the respective curve).  \label{fig:rep5}
}
\end{figure}

Studying the repetition code allows us to examine larger code distances without great computational expense.  The repetition code only corrects one type of noise.  However, we found in \fig{msc3b} that orthogonal noise contributions are largely independent.  Therefore, the repetition code provides a reasonable proxy for understanding qualitative effects of independent axes of continuous noise in a full quantum code.  \fig{rep5} shows $d=5$ repetition code performance for continuous $X$ noise with/without Pauli-$Z$ flips, as well as discrete $X$ noise for comparison.  In the repetition code, the initial codespace is irrelevant and Pauli-$Z$ flips do not cause errors, but Pauli-$Z$ flips do still induce a random walk effect since it flips the relative signs of  $\left\lvert\bar{\psi}\right\rangle$ and $\bar{X} \left\lvert\bar{\psi}\right\rangle$.
The continuous noise without the Pauli-$Z$ flips exhibits quadradic behavior from a coherent build-up of error probability.  However, when Pauli-$Z$ flips are introduced with comparable strength, there is no observed penalty for the continuous noise (it matches the discrete noise).  This is confirmation of our Sec.~\ref{sec:randwalk} analysis that indicates that this penalty is suppressed for $d>3$ codes to leading order in the noise strength.
\fig{rep5} also shows the leading order analytical results, revealing only a modest higher order correction at $\epsilon=0.15$ that does not affect our conclusions and will be even smaller for weaker noise strengths.

\begin{figure}
\includegraphics[height=6cm]{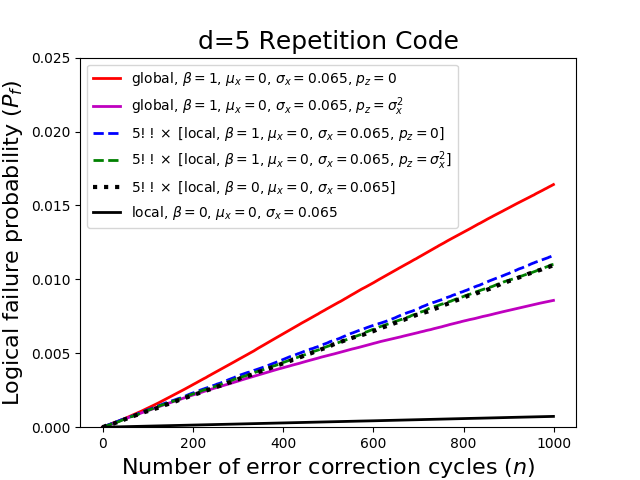}
\caption{For a $d=5$ repetition code, error correction performance due to global/local randomly sampled $X$ noise with/without temporal correlations and with/without Pauli-$Z$ flips.  Local noise cases are multiplied by $5!!$ to make a proper comparison with global noise for which a $d!!$ penalty is understood according to \secref{passive_gn} and \secref{tcgn}.  The sample uncertainty for each curve is comparable to its thickness.
\label{fig:rep5corr}
}
%We used over 200,000 trials for the global noise cases and over 800,000 trials for the local noise cases for accumulating statistical information.}
\end{figure}

Also in the $d=5$ repetition model, we examine a noise model of randomly sampled $X$ noise with temporal/spatial correlations in the extremes (perfect correlations and no correlations) shown in \fig{rep5corr}.  The performance for global noise with samples randomly drawn from a normal distribution characterized by $\mu_{\epsilon}$ and $\sigma_{\epsilon}$ is approximated in \eq{pfn_ptcgn2}.  Note that there is a $d!!$ penalty (due to the breadth of the normal distribution combined with spatial correlations) in the leading order term at $d=5$.  We confirm this in \fig{rep5corr} by comparing the performance for the global noise with the performance for the local noise multiplied by $5!!$.  With local noise, we observe a modest performance penalty when there are no $Z$ flips to induce a random walk [the second term of \eq{pfn_tcln}] which is effectively eliminated when the $Z$ flips are introduced [the second term of \eq{pfn_ptcln2}].  For the global noise, the performance penalty with $Z$ flips is much more dramatic due to the $(2d - 1)!! = 9!! = 945$ factor in the second term of \eq{pfn_tcgn_normal} compared with $d!! = 5!! = 15$ in the first term, making the penalty $63$ times larger than in the local noise case.  Again we see that introducing $Z$ flips eliminates this extra penalty from coherent noise [the second term of \eq{pfn_ptcgn2}].  In fact, here we observe a down-turn of this curve due to higher order corrections with our relatively large noise strength of $\sigma=0.065$.  This noise strength was chosen for convenience in this demonstration.  When $\sigma$ is much smaller, many more trials are required to get enough statistical samples and more steps are required to observe interesting quadratic behavior.  When $\sigma$ is much larger, higher orders influence the results but are irrelevant in a fault tolerance regime.  Our $\sigma=0.065$ results confirm our perturbative analysis and our arguments are even stronger as $\sigma$ is decreased.

%\section{Circuit model results %\label{sec:circuit_model}}
%\input{./sections/07-circuit_model.tex}

\section{Concluding Remarks \label{sec:conclusion}}
We have analyzed the effect of coherent noise on QEC.  It has long been thought that coherent noise is more harmful than independent discrete noise~\cite{Barnes2017}.  Under some circumstances, however, we have shown that these are effectively equivalent.
Although the logical failure probability under coherence noise can scale quadratically in the number of error correction cycles when performing active error correction, it has a linear form (and a proper logical failure \emph{rate}) when implementing passive error correction.
When initializing into a randomized code space, implementing passive error correction, using a code distance beyond $3$, and having a noise that is not strongly biased, we demonstrated (via perturbation theory arguments and supporting numerical simulations) the essential equivalence between a coherent noise model with strength $\delta$, as an effective Hamiltonian acting on individual qubits per error correction cycle, and discrete noise with $p = \delta^2$ as the physical gate failure probability.  

When the noise strength $\delta$ is drawn from a distribution rather than having a rigid upper bound and the noise is correlated in time, we do see a penalty beyond the standard deviation of the distribution.  This is not really an effect from coherence but rather from there being a distribution of the effective $p$ that propagates to a broadened distribution of logical failure probabilities.  In other words, the average logical failure probability is not the same as the logical failure probability for the average $p$.

In this work, we have not addressed the effects of coherent multi-qubit (correlated) errors during the syndrome extraction process. Such coherent rotations may add constructively (or destructively) in between error correction cycles for a legitimate penalty due the coherence of the noise. If one can bound
an effective noise strength over an entire error correction cycle as weight-1 Hamiltonian terms, then our theory may be applied to that bounded noise strength and thereby bound the logical error rate.

We would like to acknowledge Andrew Landahl for helpful discussions and Richard Muller who developed the vector state simulator that we used to generate results.  Sandia National Laboratories is a multi-mission laboratory managed and operated by National Technology \& Engineering Solutions of Sandia, LLC (NTESS), a wholly owned subsidiary of Honeywell International Inc., for the U.S. Department of Energy’s National Nuclear Security Administration (DOE/NNSA) under contract DE-NA0003525. This written work is authored by an employee of NTESS. The employee, not NTESS, owns the right, title and interest in and to the written work and is responsible for its contents. Any subjective views or opinions that might be expressed in the written work do not necessarily represent the views of the U.S. Government. The publisher acknowledges that the U.S. Government retains a non-exclusive, paid-up, irrevocable, world-wide license to publish or reproduce the published form of this written work or allow others to do so, for U.S. Government purposes. The DOE will provide public access to results of federally sponsored research in accordance with the DOE Public Access Plan.

\bibliographystyle{unsrt}
\bibliography{bibliography}

\end{document}